\documentclass[reprint,amsmath,amssymb,aps,superscriptaddress]{revtex4-1}
\usepackage{graphicx}
\usepackage{dcolumn}
\usepackage{bm}
\usepackage{float}
\usepackage{color}

\begin{document}

% EXPLAIN CBM, VBM

\title{Electronic structure of monolayer antimonene nanoribbons under out-of-plane and transverse bias}

\author{Edo van Veen}
\affiliation{Institute for Molecules and Materials, Radboud University, NL-6525 AJ Nijmegen, The
Netherlands}
\author{Jin Yu}
\affiliation{Beijing Computational Science Research Center, Beijing 100094, China}
\affiliation{Institute for Molecules and Materials, Radboud University, NL-6525 AJ Nijmegen, The
Netherlands}
\author{Mikhail I. Katsnelson}
\affiliation{Institute for Molecules and Materials, Radboud University, NL-6525 AJ Nijmegen, The
Netherlands}
\author{Rafael Rold\'an}
\affiliation{Materials Science Factory. Instituto de Ciencia de Materiales de Madrid (ICMM), Consejo Superior de Investigaciones Cient\'{i}ficas (CSIC), Cantoblanco E28049
Madrid, Spain}
\author{Shengjun Yuan}
\email{s.yuan@whu.edu.cn}
\affiliation{School of Physics and Technology, Wuhan University, Wuhan 430072, China}

\date{\today}

\begin{abstract}
A systematic study of the electronic properties of single layer Sb (antimonene) nanoribbons is presented. By using a 6-orbital tight-binding Hamiltonian, we study the electronic band structure of finite ribbons with zigzag or armchair termination. We show that there is good agreement between {\it ab initio} calculations and the tight-binding model. We study how the size of the gap can be controlled by applying an external bias potential. An electric field applied perpendicular to the antimonene layer is found to increase the band gap, while a transverse bias potential leads to a position dependent reduction of the band gap. Both kinds of bias potential break inversion symmetry of the crystal. This, together with the strong intrinsic spin-orbit coupling of antimonene, leads to spin-splitting of the valence band states.
%\begin{description}
%\end{description}
\end{abstract}

\pacs{PACS}

\maketitle

\section{Introduction}

Two dimensional (2D) materials \cite{Novoselov_PNAS_2005}, such as graphene, transition metal dichalcogenides and hexagonal boron nitride, are attracting tremendous interest due to their unique electronic, optical and mechanical properties, remarkably different from their three-dimensional counterparts \cite{Roldan_CSR_2017}. Recently, the family of 2D materials derived from the group-VA layered crystals (P, As, Sb, Bi) has been the focus of great attention \cite{yu2018tunable, Zhang_CSR_2018}, black phosphorus being the most well studied among them. In 2015 Zhang {\it et al.} predicted that, contrary to bulk antimony which is a semimetal, single-layer Sb (antimonene) is an indirect band gap semiconductor \cite{Zhang_AC_2015}. Soon after, it was demonstrated that atomically thin antimonene can be obtained by different means, including van der Waals epitaxy \cite{Ji_NC_2016}, micromechanical exfoliation \cite{Ares_AM_2016}, liquid phase exfoliation \cite{Gibaja_ACIE_2016}, molecular beam epitaxy \cite{Wu_AM_2017} or electrochemical exfoliation \cite{Lu_AOM_2017}. Theoretical calculations have studied in detail the electronic properties of this material \cite{Wang_ACSAMI_2015,Akturk_PRB_2015,Singh_JMCC_2016,Pizzi_NC_2016,Xu_AP_2017}. Strong spin-orbit coupling was also reported, with a coupling strength of $\lambda\approx 0.34$~eV \cite{Rudenko_PRB_2017}. {\it Ab initio} quantum transport calculations have shown that antimonene field effect transistors (FETs) can satisfy both the low power and high performance requirements for usage in nanoscale electronic and optoelectronic devices \cite{Wang_CM_2017}. Previous experience with graphene and other 2D materials has further motivated theoretical studies of the electronic properties of nanoribbons of group-VA semiconductors  \cite{Guo_JPCC_2014,Taghizadeh_PRB_2015,wang2015electronic,Grujic_PRB_2016,Song_npjQM_2017}. Recently, experimental fabrication of antimonene nanoribbons has been reported \cite{Tsai_CC_2016}, demonstrating band gap opening due to quantum confinement. 

In this work we study the band structure and electronic properties of Sb nanoribbons in the presence of out-of-plane and in-plane electric fields. We find that edge states are present in nanoribbons with both zigzag and armchair termination. We find good agreement between {\it ab initio} numerical simulations and tight-binding calculations. We further demonstrate that the size of the band gap can be controlled by the presence of an external bias field. Application of a bias field breaks inversion symmetry which, together with the strong spin-orbit coupling in antimonene, leads to splitting of the valence band edges, with the corresponding spin-valley coupling due to Rashba effect. 

\begin{figure}[t]
\includegraphics[width=0.45\linewidth]{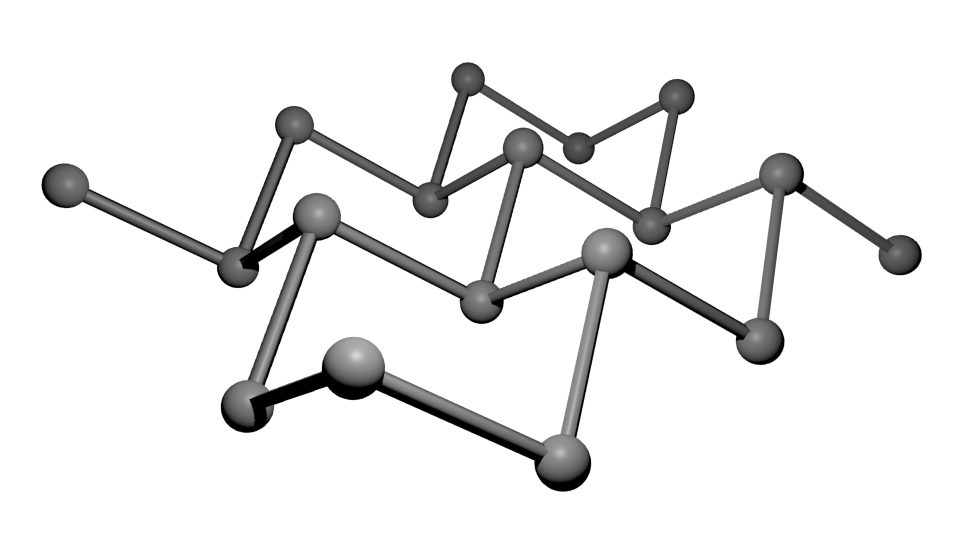}
\includegraphics[width=0.25\linewidth]{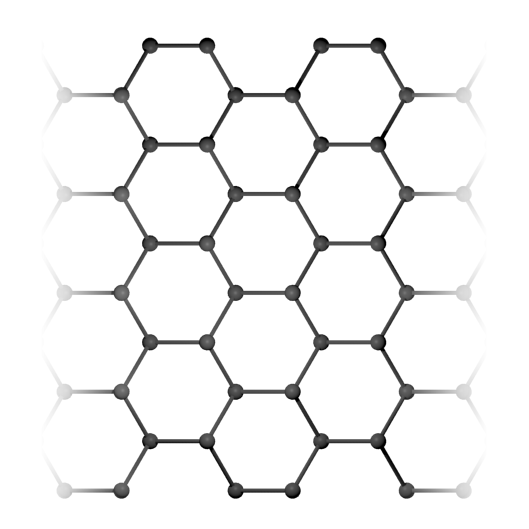}
\includegraphics[width=0.25\linewidth]{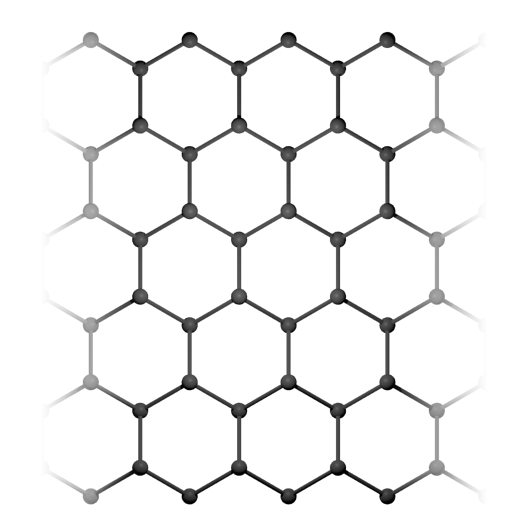}
\includegraphics[width=\linewidth]{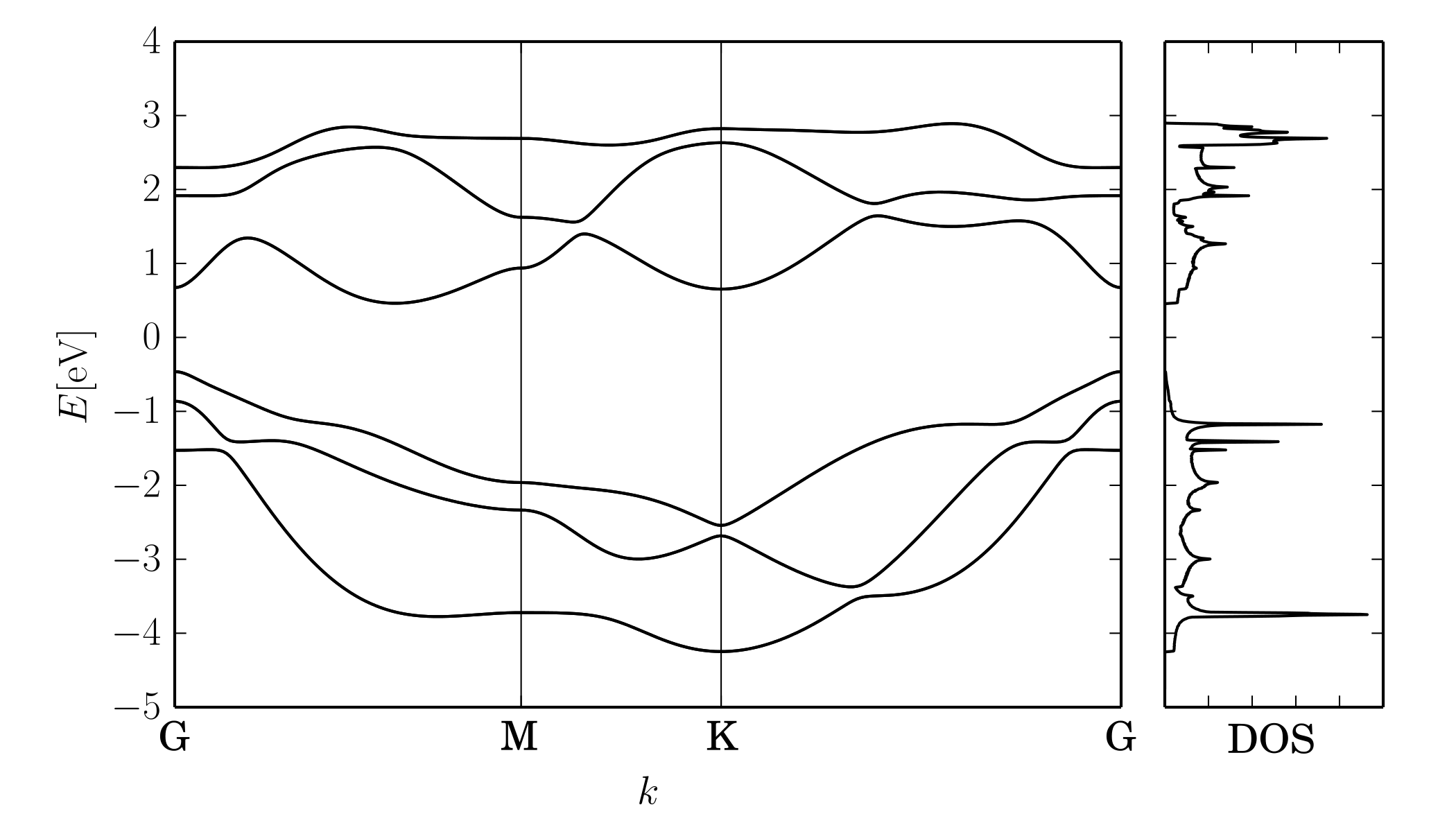}
\caption{Top: the buckled honeycomb lattice structure of antimony (left), an armchair SbNR (middle) and a zigzag SbNR (right). Bottom: band structure and DOS for pristine Sb calculated with the tight-binding Hamiltonian (\ref{Eq:Hamiltonian_Real}), with the hopping parameters given in Table \ref{Tab:Hoppings}.}
\label{Fig:pristine}
\end{figure}

The paper is organized as follows. In Sec. \ref{Sec:Model} we describe the tight-binding model and the details of the calculations. We also show results for unbiased nanoribbons. In Sec. \ref{Sec:Out-of-plane} we study the effect of a perpendicular electric field on the electronic properties and the band structure of Sb-nanoribbons, and the effect of an in-plane bias field is studied in Sec. \ref{Sec:In-plane}. Our main results are summarised in Sec. \ref{Sec:Conclusions}.

\section{Model and method}\label{Sec:Model}

\begin{figure}[t]
\includegraphics[width=0.35\textwidth, angle=0]{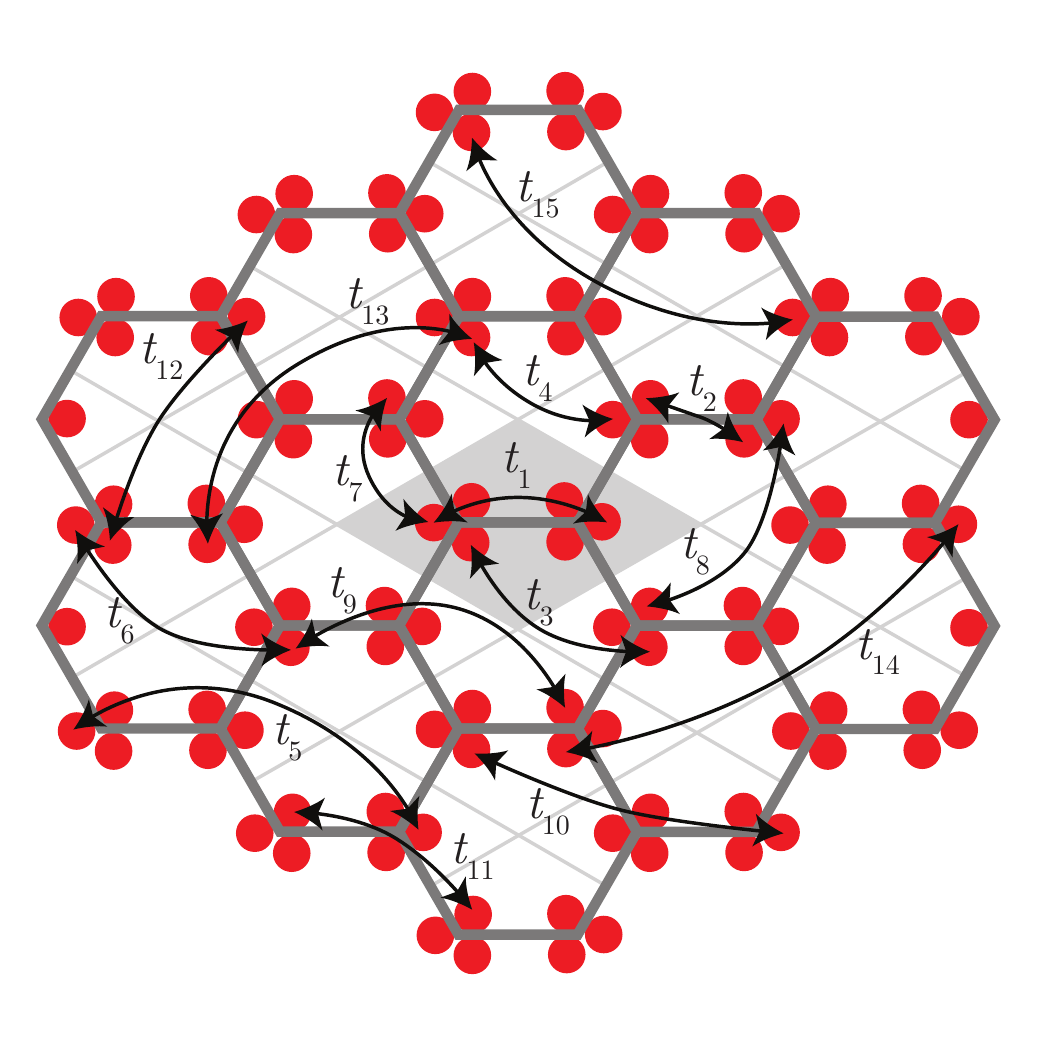}
\caption{Top view of antimonene crystal structure with the hopping parameters ($t_i$)
included in the TB model. Their corresponding values are given in \mbox{Table \ref{Tab:Hoppings}}. The red circles represent $p$-orbitals.}
\label{Fig:Hoppings}
\end{figure}

\noindent 
Single layer antimonene consists on a buckled honeycomb lattice of Sb atoms (Fig. \ref{Fig:pristine}), with the two sublattices vertically displaced by $b=1.65$~\AA, and with an in-plane lattice constant of $a=4.12$~\AA. The relevant energy bands of the electronic structure, including SOC effects, are very well captured by a 6-orbitals tight-binding Hamiltonian developed by Rudenko {\it et al.} \cite{Rudenko_PRB_2017}, which includes the 3 $p$-orbitals of each of the two Sb atoms of the unit cell:
\begin{eqnarray}\label{Eq:Hamiltonian_Real}
	 H = & \sum_{m} \sum_{i} \sum_{\sigma} \epsilon^{\phantom{\dagger}}_{mi\sigma} 
		c^{\dagger}_{mi\sigma} c^{\phantom{\dagger}}_{mi\sigma}\\ \nonumber
	 	& + \sum_{mn} \sum_{ij} \sum_{\sigma} t^{\phantom{\dagger}}_{mi \sigma; nj \sigma} 
		c^{\dagger}_{mi\sigma} c^{\phantom{\dagger}}_{nj\sigma}\\ \nonumber
		& + \sum_{mn} \sum_{i} \sum_{\sigma \sigma'} h^{\phantom{\dagger}}_{mi \sigma; ni \sigma'} 
		c^{\dagger}_{mi\sigma} c^{\phantom{\dagger}}_{ni\sigma'}
\end{eqnarray}
where $m, n$ run over orbitals, $i, j$ run over sites and $\sigma, \sigma'$ run over spins; $c^{\dagger}_{mi\sigma}$ ($c_{mi\sigma}$) is the creation (annihilation) operator on orbital $m$ at site $i$ with spin $\sigma$. The parameters $\epsilon_{mi\sigma}$ account for on-site potentials, $t_{mi \sigma; nj \sigma}$ are inter-orbital hopping terms, and intra-atomic SOC is accounted by  $h_{mi \sigma; ni \sigma'}$. The intra-atomic SOC constant is $\lambda=0.34$~eV and the hopping parameters are given in Table \ref{Tab:Hoppings} \cite{Rudenko_PRB_2017} and schematically shown in Fig. \ref{Fig:Hoppings}. 
The corresponding density of states (DOS) is calculated from
\begin{equation}
	 D(E) = \frac{1}{2 \pi} \sum_{n} \int_{BZ} \delta(E-E_n(k)) dk,
\end{equation}
where $n$ labels the different energy bands. The band structure and DOS obtained with this model for bulk antimonene are shown in Fig. \ref{Fig:pristine}. Single layer antimonene is an indirect gap semiconductor with a band gap of 0.92 eV. The edge of the valence band is located at the $\Gamma$ point of the BZ, with main contributions from $p_x$ and $p_y$ orbitals, while the edge of the conduction band is placed at a non high-symmetry point of the BZ, with relevant contributions from all 3 $p$-orbitals of Sb. 

   \begin{table}[b]
    \centering
    \caption[Bset]{Hopping amplitudes $t_i$ (in eV) entering in the TB Hamiltonian Eq.~(\ref{Eq:Hamiltonian_Real}), as obtained in \cite{Rudenko_PRB_2017}. $d$ denotes the 
                   distance between the lattice sites on which the interacting orbitals are centered.
                   $N_c$ is the corresponding coordination number. The hoppings are schematically shown in Fig.~\ref{Fig:Hoppings}.}
 \begin{tabular}{cccc|cccc|cccc}
      \hline
      \hline
 $i$&  $t_i$(eV) &  $d$(\AA) &  $N_c$  & $i$&  $t_i$(eV) &  $d$(\AA) &  $N_c$ &  $i$ &  $t_i$(eV) &  $d$(\AA)  &  $N_c$ \\
     \hline
   1&    -2.09    &    2.89    &    1    &  6 &     0.21    &    4.12    &    1   &      11 &   -0.06    &    4.12    &   2  \\
   2&     0.47    &    2.89    &    2    &  7 &     0.08    &    2.89    &    2   &      12 &   -0.06    &    5.03    &   1  \\
   3&     0.18    &    4.12    &    4    &  8 &    -0.07    &    5.03    &    2   &      13 &   -0.03    &    6.50    &   2  \\
   4&    -0.50    &    4.12    &    1    &  9 &     0.07    &    6.50    &    2   &      14 &   -0.04    &    8.24    &   1  \\
   5&    -0.11    &    6.50    &    2    &  10&     0.07    &    6.50    &    2   &      15 &   -0.03    &    8.24    &   1  \\
      \hline                  
      \hline
    \end{tabular}
\label{Tab:Hoppings}
    \end{table}

\begin{figure}[t]
\includegraphics[width=\linewidth]{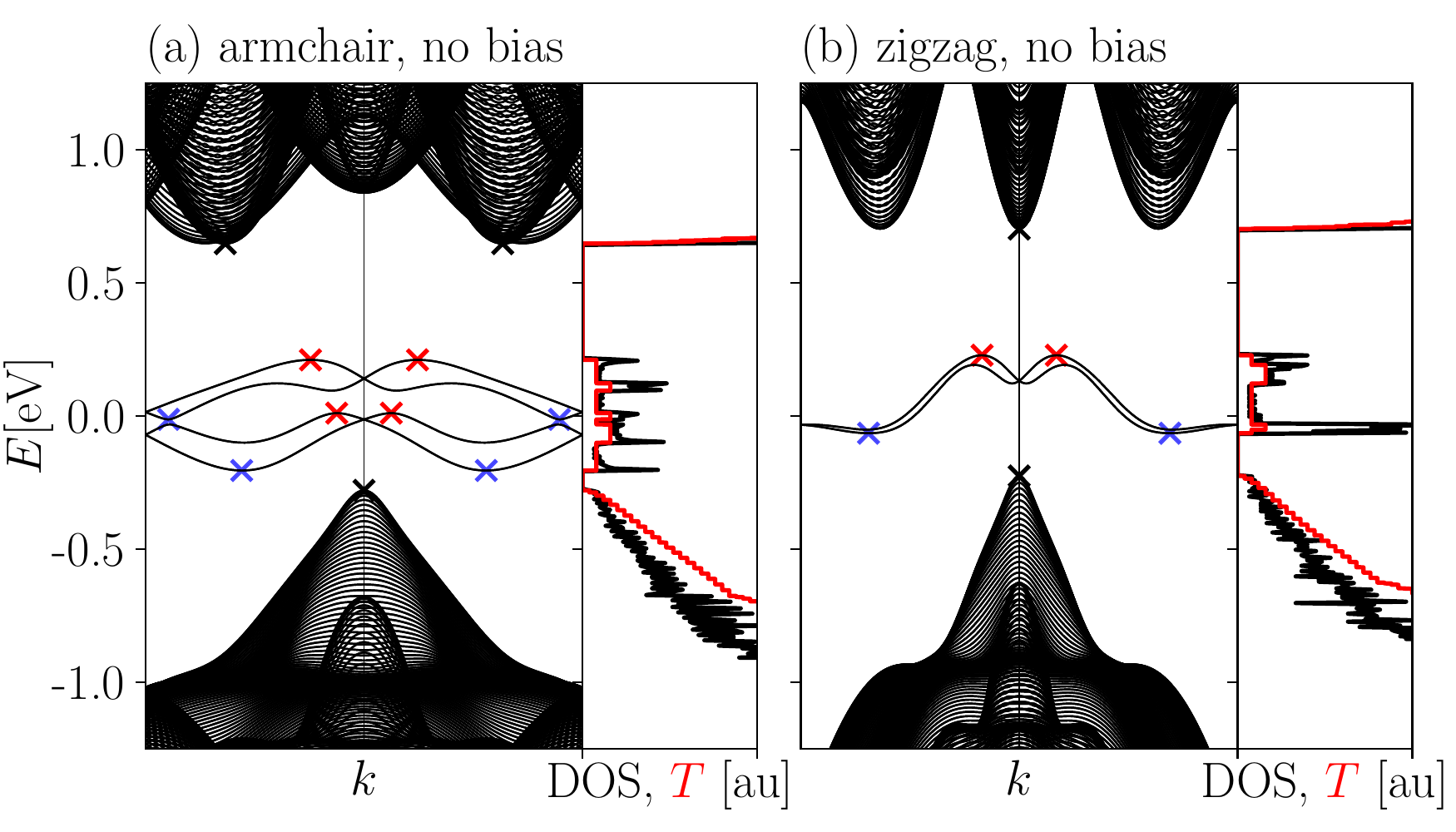}
\caption{Band structure, DOS and transmission for (a) an armchair ribbon of width 41 nm and (b) a zigzag ribbon of width 36 nm. The conduction and valence band edges are indicated with black crosses, the edge band maxima are red and the edge band minima are blue. The midgap bands correspond to edge states.}
\label{Fig:Bands_Ribbons}
\end{figure}

\begin{figure}[t]
\includegraphics[width=0.49\linewidth]{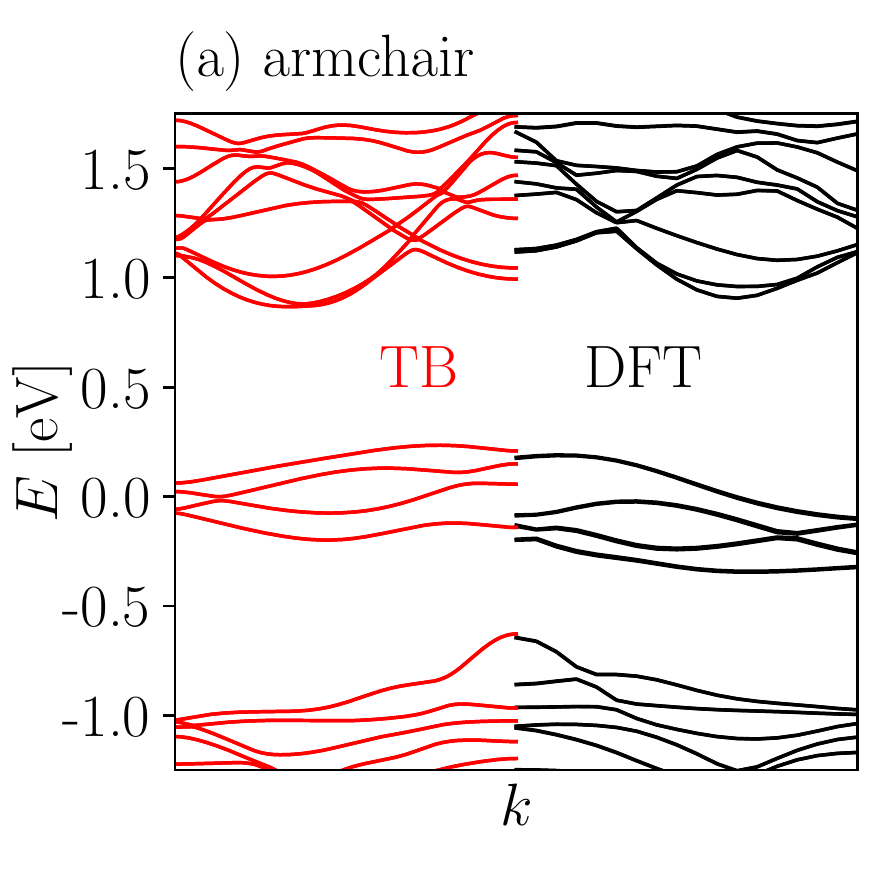}
\includegraphics[width=0.49\linewidth]{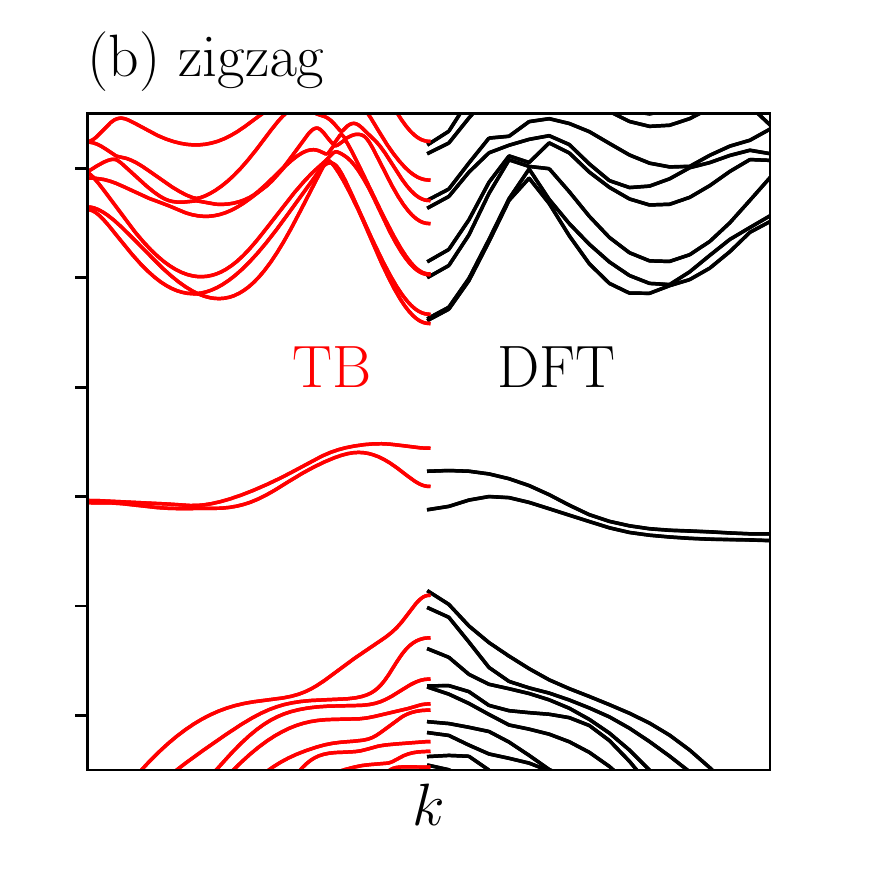}
\caption{Comparison of the band structure of nanoribbons as obtained from tight-binding and DFT methods: (a) a 2.3 nm width ribbon with armchair termination, and (b) a 2.9 nm width ribbon with zigzag termination. Red corresponds to TB method and black to DFT calculations.}
\label{Fig:compare-TB-DFT}
\end{figure}

\begin{figure}[t]
\includegraphics[width=0.49\linewidth]{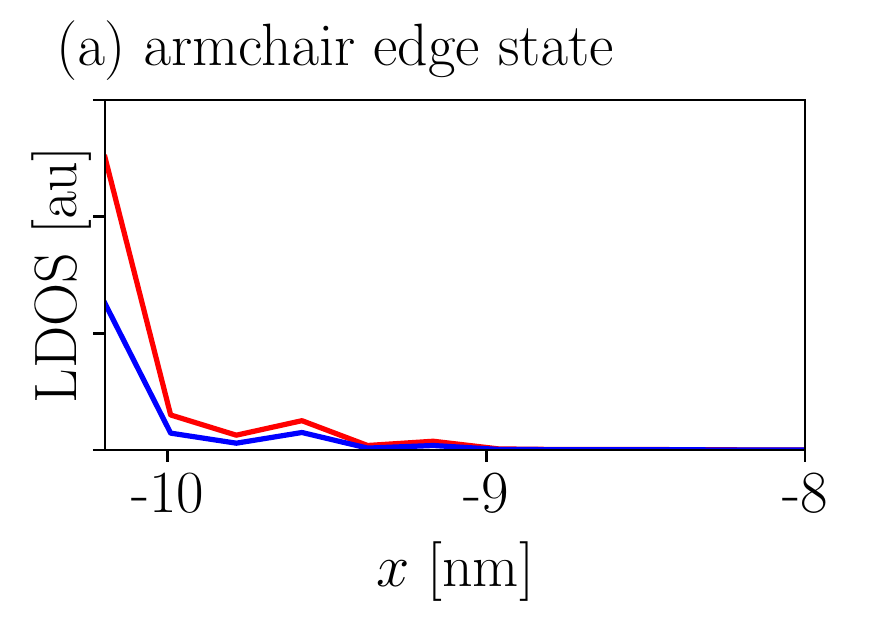}
\includegraphics[width=0.49\linewidth]{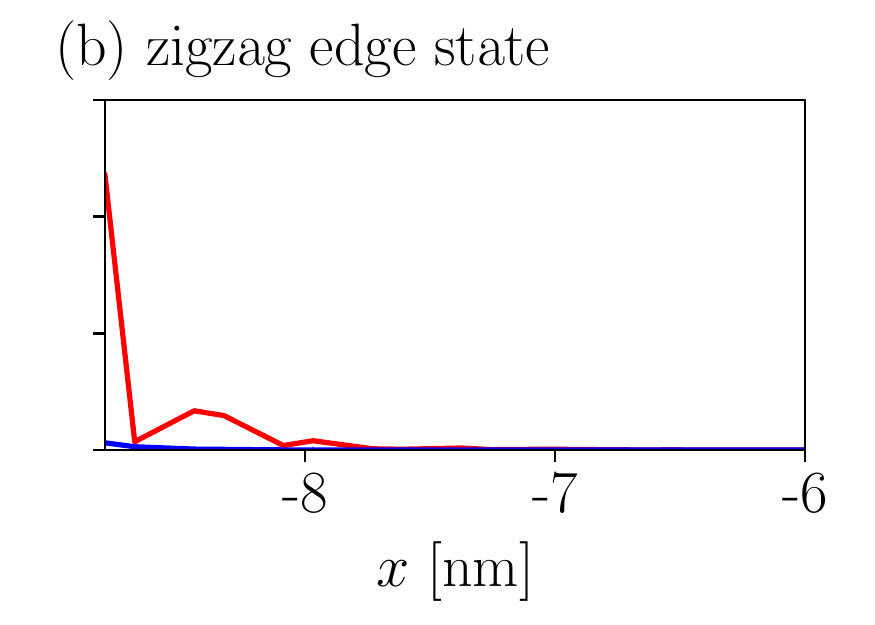}
\caption{LDOS of edge states at $k=\frac{\pi}{2 W}$ for (a) an armchair ribbon of width 20 nm and (b) a zigzag ribbon of width 18 nm. The red and blue lines correspond to spin up and down, respectively. Notice that these states are degenerate with the corresponding states for $k=-\frac{\pi}{2 W}$, which have opposite spin.}
\label{Fig:wfs_edge}
\end{figure}

Since we are interested in electronic properties of semi-infinite ribbons, the momentum parallel to the infinite edge is a good quantum number and we can Fourier-transform Hamiltonian (\ref{Eq:Hamiltonian_Real}) along that direction.
The band structure of finite nanoribbons (we impose periodic boundary conditions along the direction parallel to the edge) with zigzag and armchair termination are shown in Fig. \ref{Fig:Bands_Ribbons}. Firstly, the finite width of the antimonene ribbon leads to a reconstruction of the band structure with the formation of electronic bands composed by the accumulation of $N$ subbands, where $N$ is the number of unit cells along the width of the ribbon. Secondly, midgap edge states appear in both armchair and zigzag nanoribbons (Fig. \ref{Fig:Bands_Ribbons}), originating from the unsaturated bond on the edge of the ribbon.
This is different from graphene and black phosphorus ribbons, for which edge states are absent for armchair termination \cite{Brey_PRB_2006,Grujic_PRB_2016}. The energy bands associated to the edge states are flat and weakly dispersing, leading to prominent peaks in the DOS associated to saddle points in the band structure.

In this work, we only consider chemically unsaturated edges, i.e., we do not take edge chemistry into account. Attaching different atoms to the edge could significantly alter the electronic structure around the Fermi level \cite{gunlycke2007altering}.

To check whether our bulk TB model agrees with {\it ab initio} calculations also for finite ribbons, we performed first-principles calculations on the electronic structure of antimony nanoribbons, including SOC, using the Vienna {\it ab initio} simulation package (VASP) \cite{kresse1996g, kresse1996g2}. Electron exchange and correlation interactions were described using the Perdew-Burke-Ernzerhof (PBE) pseudopotentials within the projector augmented-wave method \cite{blochl1994pe}. The Brillouin zone sampling was done using a 35*1*1 Monkhorst-Pack grid for static calculation. The atomic structure of the nanoribbons are obtained from the 2D nanosheet without structure relaxation, and the vacuum region between two adjacent images is set to be 100 \AA. The results are shown in Fig. \ref{Fig:compare-TB-DFT}, in comparison to tight-binding calculations. We can see that the agreement between the two methods is reasonable. Apart from some slight shifts in energies,  the contours of the conduction band minimum (CBM), valence band maximum (VBM) and edge states in the TB model are very similar to the DFT result. We notice that previous first-principles calculations for narrow nanoribbons, of up to $\sim$3.4~nm, predicted a direct band gap for zigzag termination \cite{wang2015electronic}, which is also in agreement with our own TB calculations. By systematically studying the evolution of the bandgap with nanoribbon size, we conclude that the two secondary CBMs around $k = \pm 0.63 \frac{\pi}{W}$ get closer to the the CBM at $k = 0$ for increasing size. For a ribbon width of 175 nm, the difference between their energy values is only on the order of $10^{-5}$ eV.

The midgap states are highly localized at the edges, as can be seen in Fig. \ref{Fig:wfs_edge}. For zero bias, each $\bf k$ presents two degenerate states with opposite spin in opposite sides of the ribbon. Since time-reversal symmetry must be preserved, the spin polarization of one edge associated to one state of a given wave-vector $\bf k$, is compensated by the opposite spin polarization of the degenerate state with momentum $-\bf k$.

In the following, we use Landauer theory \cite{landauer1957spatial} to calculate the electronic transmission in the scattering-free limit, which is obtained by counting modes:
\begin{equation}
T(E) = \sum_k N_k (E),
\end{equation}
where $N_k(E)$ is the number of bands that cross the energy $E$ for a given wave-vector $k$. The results for $T(E)$ for each termination are shown in Fig. \ref{Fig:Bands_Ribbons} (solid red lines), together with the DOS (solid black lines). 

The main difference between the transmission in zigzag and armchair nanoribbons occurs for energies within the bulk bandgap. At these energies $T(E)$ is dominated by edge states which, as we have seen, are different for zigzag and armchair terminations.  As the ribbon width increases, the transmission function corresponding to the bulk states increases, accompanied by  a reduction of the energy gap, while the transmission of the edge states remains the same. 

We have further calculated the effective mass of electrons $m_e^*$ and holes  $m_h^*$ from the nanoribbon band structure (table \ref{Tab:EffectiveMasses}). Electrons are heavier than holes for both edge terminations. We also find that carriers in zigzag nanoribbons are expected to have lower effective masses than in armchair nanoribbons. These results can be useful for calculations based on low energy ${\bf k}\cdot{\bf p}$ analytical models. 
\iffalse
\begin{equation}
m^* =\hbar^2\left(  \frac{d^2E}{dk^2} \right)^{-1}.
\end{equation}
\fi

   \begin{table}[!bt]
    \centering
    \caption[Bset]{Effective masses for antimonene nanoribbons.}
 \begin{tabular}{c|c|c}
      \hline
      \hline
 edge &  $m^*_e$ ($m_0$) &  $m^*_h$ ($m_0$) \\
     \hline
   armchair &    0.2 & 0.13  \\
   zigzag &    0.13 & 0.09  \\
      \hline                  
      \hline
    \end{tabular}
\label{Tab:EffectiveMasses}
    \end{table}
    
    \begin{figure}[t]
\includegraphics[width=\linewidth]{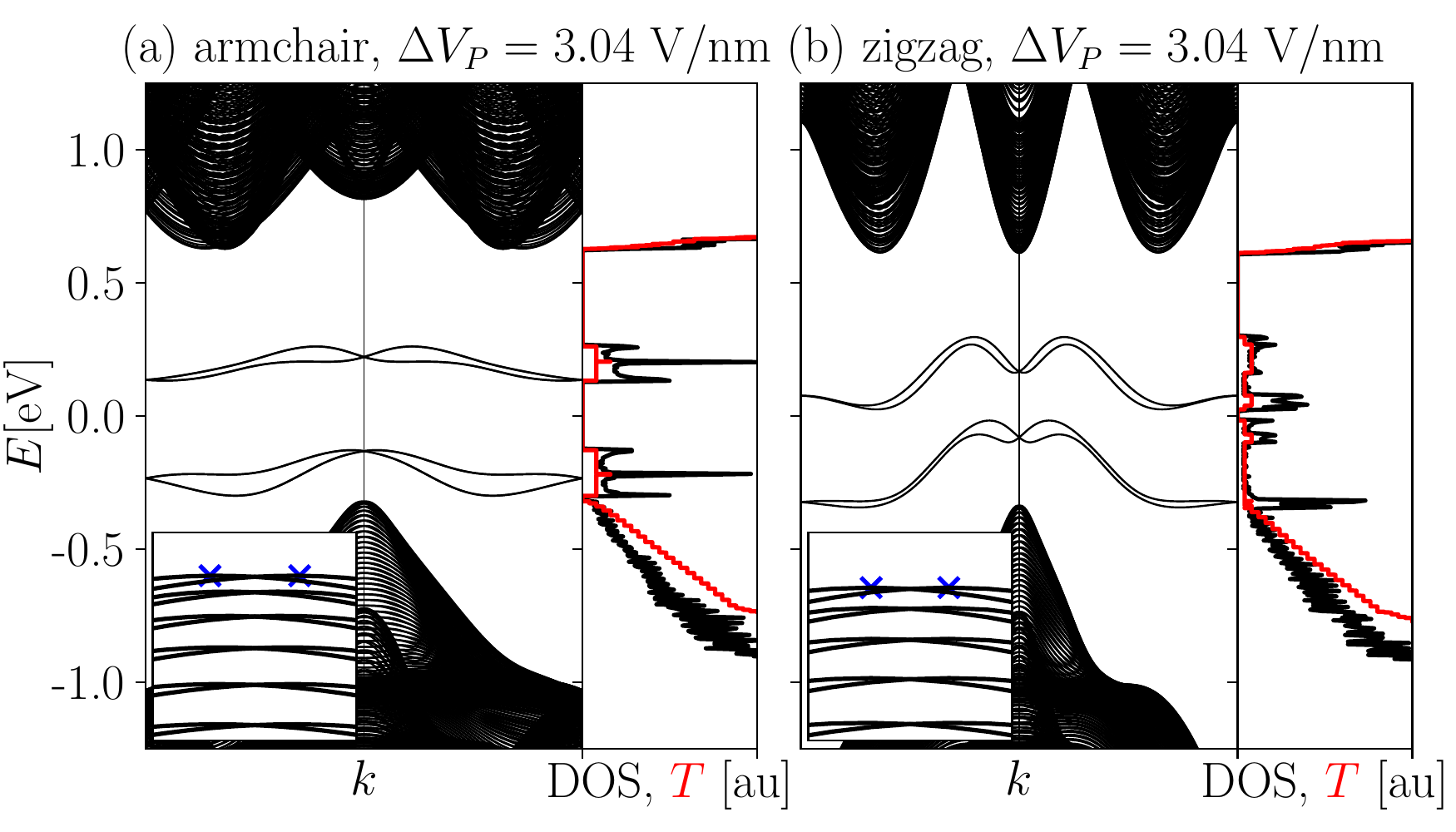}
\caption{Top: band structure, DOS and transmission with $\Delta V_P= 3.04$ V/nm for (a) an armchair ribbon of width 41 nm and (b) a zigzag ribbon of width 36 nm. Insets: close up of the edge of the valence band, with blue crosses to indicate the maxima.}
\label{Fig:Out-of-plane-DOS}
\end{figure}

\begin{figure}[t]
\includegraphics[width=0.49\linewidth]{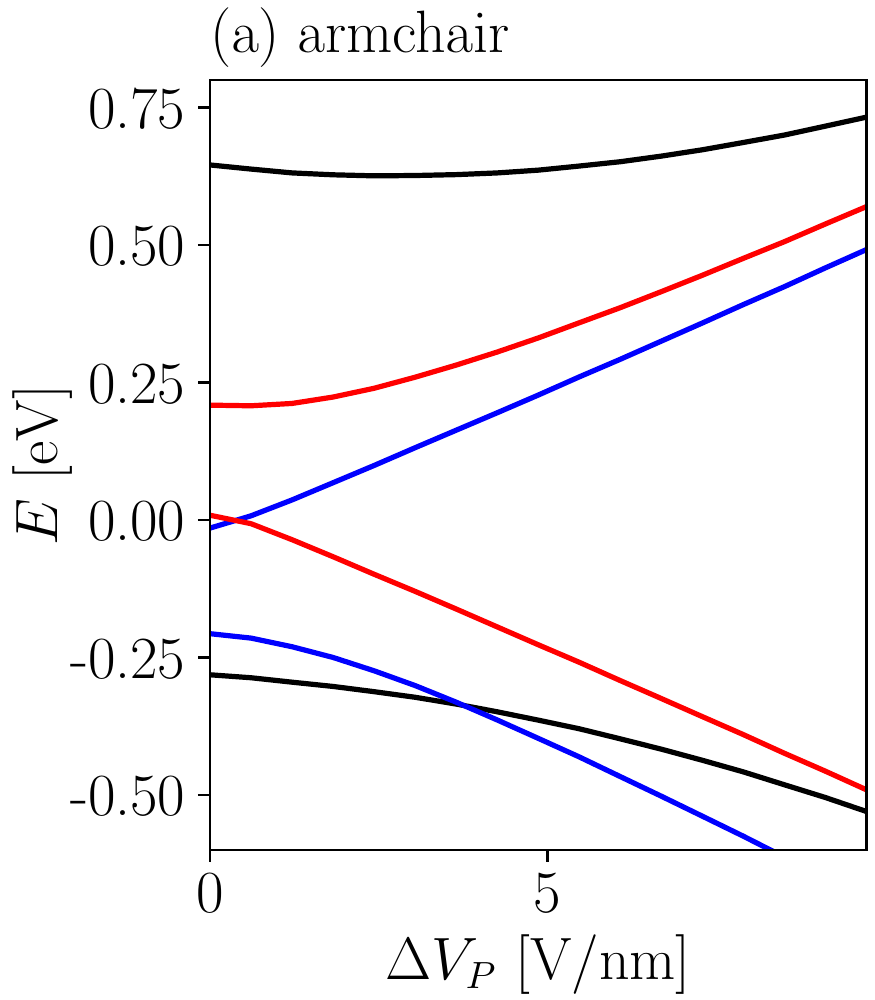}
\includegraphics[width=0.49\linewidth]{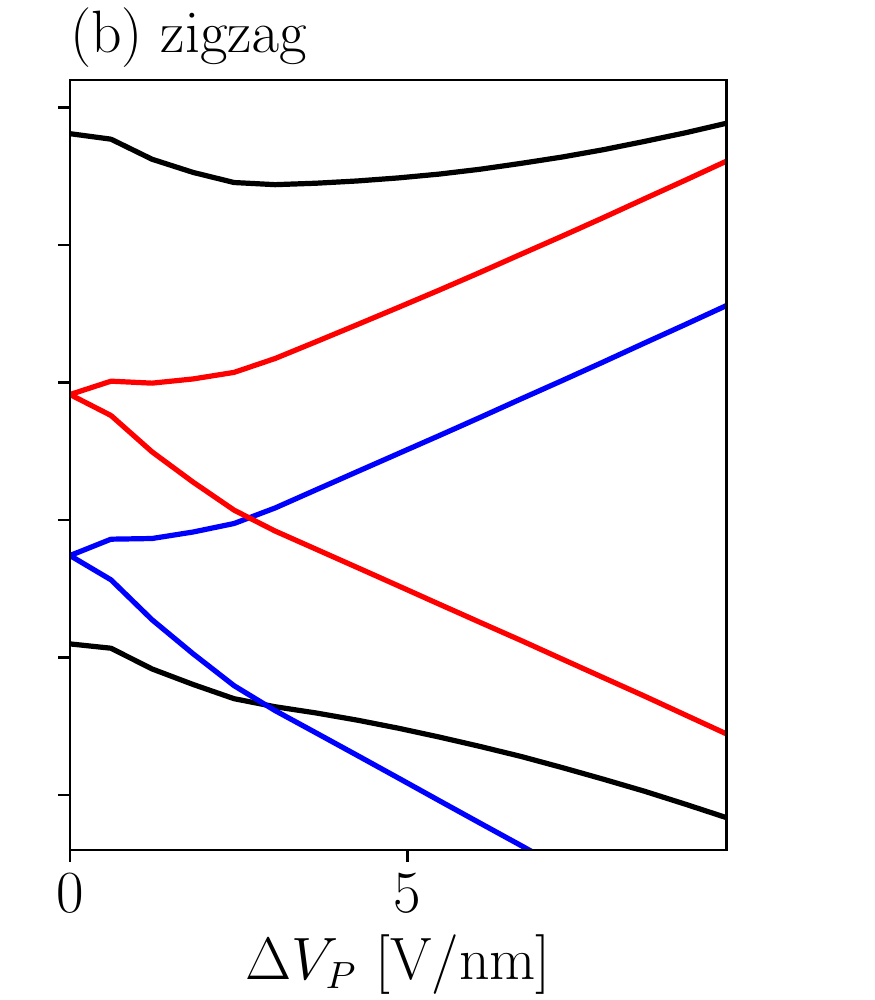}
\caption{Band edges as function of $\Delta V_P$ for an armchair ribbon of width 41 nm (left) and a zigzag ribbon of width 36 nm (right). Black lines correspond to the conduction and valence band edges, and red (blue) corresponds to the maxima (minima) of the edge states, corresponding to the crosses in Figure \ref{Fig:Bands_Ribbons}.}
\label{Fig:Out-of-plane-Gap}
\end{figure}

\section{Out-of-plane bias}\label{Sec:Out-of-plane}
Application of external gate voltages is a powerful tool to control and tune the electronic and optical properties of layered 2D materials \cite{Roldan_NP_2017}. In this section we study the effect of a perpendicular bias voltage on antimonene. 
Since monolayer Sb is buckled, the application of an electric field perpendicular to the sample leads to a potential difference between atoms in different planes. Therefore we introduce an out-of-plane bias $\Delta V_P$ (without considering screening) by setting the on-site potential on the two sublattices in Hamiltonian (\ref{Eq:Hamiltonian_Real}) to different values:
\begin{equation}
	\epsilon_{mi\sigma} = \Delta V_P \times z_i,
\end{equation}
where $z_i$ is the $z$-coordinate of site $i$ in the buckled structure, which is plus or minus $0.82$ \AA\ on sublattice A or B, respectively. Our results for the zigzag and armchair nanoribbon band structure, with the corresponding DOS and electronic transmission, are given in Fig. \ref{Fig:Out-of-plane-DOS}. First, we notice that for both types of ribbon, we obtain a bandgap widening under the application of the electric field. The evolution with the applied bias of the valence and conduction bands, as well as the edge states, are shown in Fig. \ref{Fig:Out-of-plane-Gap}. Opening of the bandgap with electric field was also predicted for single-layer black phosphorus \cite{Yuan_PRB_2016}. Interestingly, application of a bias voltage breaks inversion symmetry (sublattices A and B are no longer equivalent). This, together with the strong spin-orbit coupling leads, due to Rashba effect, to splitting of the edge states, and of the valence and conduction bands. Notice that, because of the latter, the zigzag ribbon band gap becomes indirect when a bias is applied (see insets of Fig. \ref{Fig:Out-of-plane-DOS} for a close-up of the valence band edge). Application of a perpendicular bias field opens, therefore, the possibility to dynamically tuning the Rashba energy \cite{Ast_PRB_2008}, or to study unconventional transport properties associated to entanglement between spin and charge degrees of freedom \cite{Brosco_PRL_2016}.

\begin{figure}[t]
\includegraphics[width=0.49\linewidth]{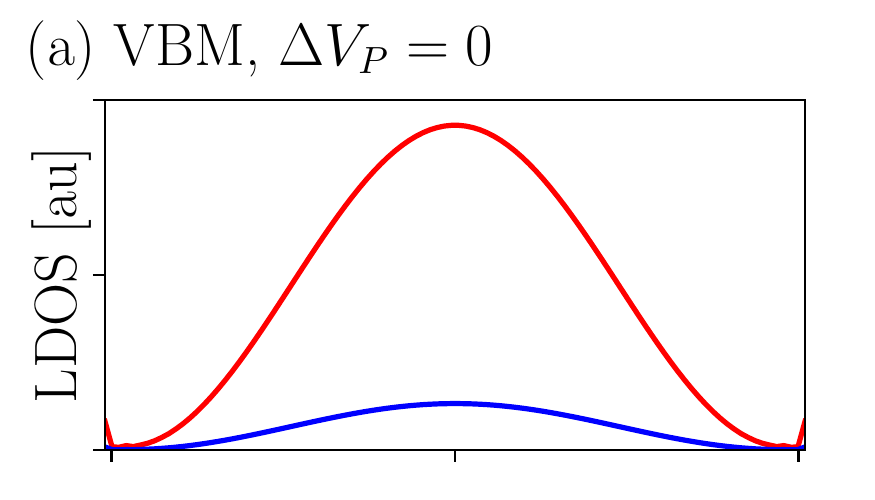} 
\includegraphics[width=0.49\linewidth]{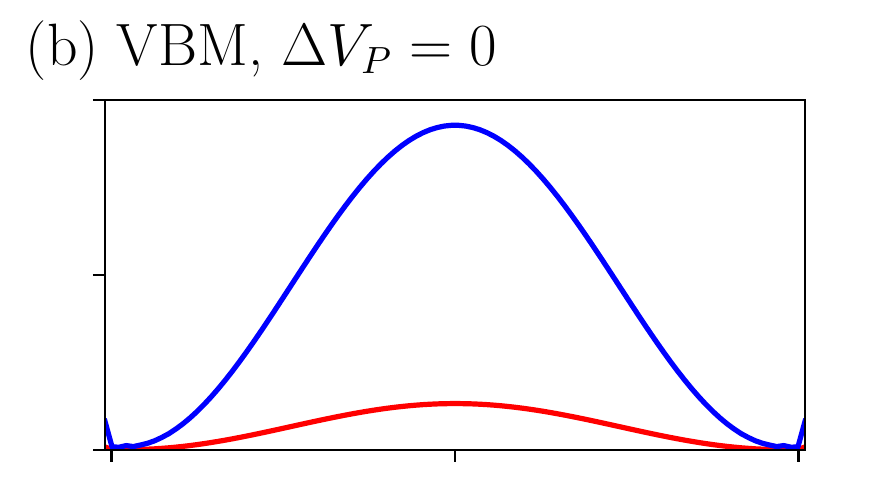} \\
\includegraphics[width=0.49\linewidth]{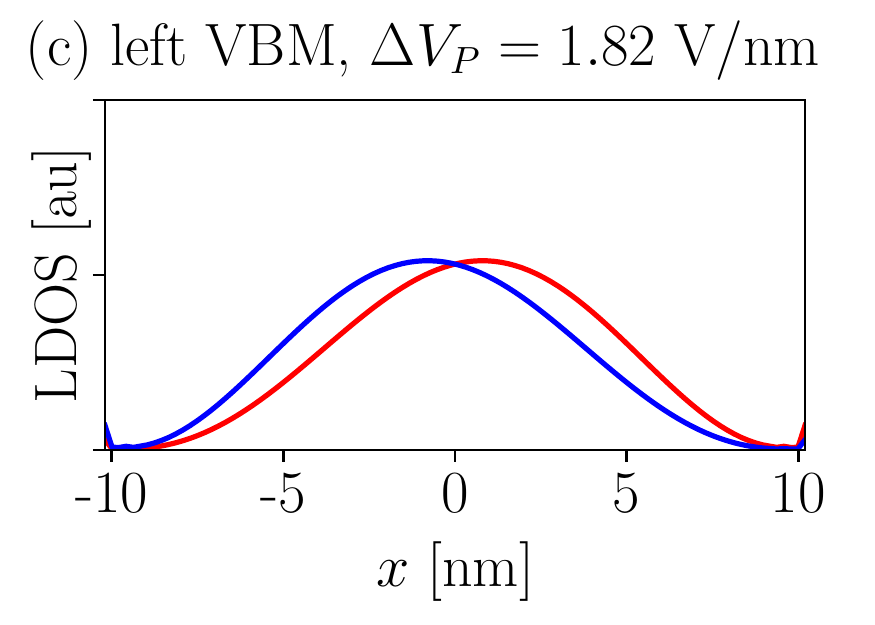}
\includegraphics[width=0.49\linewidth]{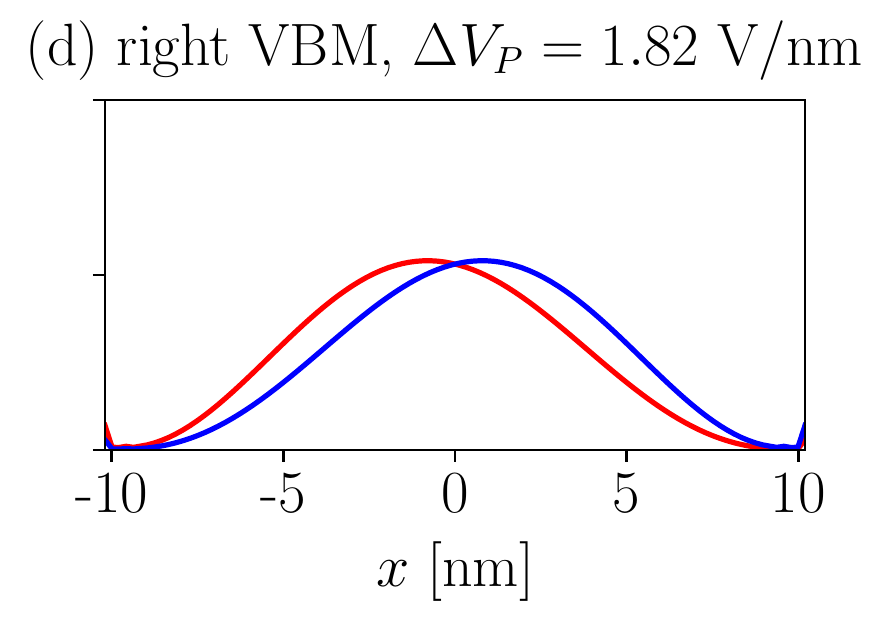}
\caption{LDOS of the valence band maxima for an armchair ribbon of width 20 nm, under out-of-plane bias. The red and blue lines correspond to spin up and down, respectively.}
\label{Fig:Out-of-Plane-LDOS}
\end{figure}

The local distribution of the eigenstates can be investigated by calculating the Local Density of States (LDOS) \cite{Rostami_JPCM_2016}, which is the probability amplitude as a function of location, in the transverse direction of the ribbon. Our results for the LDOS corresponding to the valence band are shown in Fig. \ref{Fig:Out-of-Plane-LDOS}. First, we notice that in the absence of a bias field, the maximum of the valence band is doubly degenerate (due to spin). The local distribution of the VB states is maximum at the center of the ribbon, and decays as we approach the ribbon edges (panels (a) and (b)). The situation is different when an out-of-plane electric field is applied: as discussed above, due to Rashba coupling the edge of the VB is split into two maxima around $\Gamma$ (see insets of Fig. \ref{Fig:Out-of-plane-DOS}) and the spin degeneracy is broken. The consequence of this on the LDOS is seen in Fig. \ref{Fig:Out-of-Plane-LDOS} (c) and (d). The states corresponding to the left maximum (panel (c)) present major contribution of spin down (up) at the left (right) side of the center of the ribbon. Of course, since time-reversal symmetry must be conserved, the opposite happens for the states associated to the right VBM (panel (d)).

For the armchair ribbon, for nonzero out-of-plane bias, the two lower midgap bands move down and the two upper bands move up. For the zigzag case, however, the two midgap bands that were originally doubly degenerate, split into two pairs of non-degenerate bands. The edge states on one side of the ribbon move up in energy, while the states on the other side move down. This is due to the fact that the sites on one edge of the ribbon have $z$-coordinate of $+0.82$ \AA\ and on the other edge $-0.82$ \AA, because of the buckled structure.

\section{In-plane transverse bias}\label{Sec:In-plane}

\begin{figure}[t]
\includegraphics[width=\linewidth]{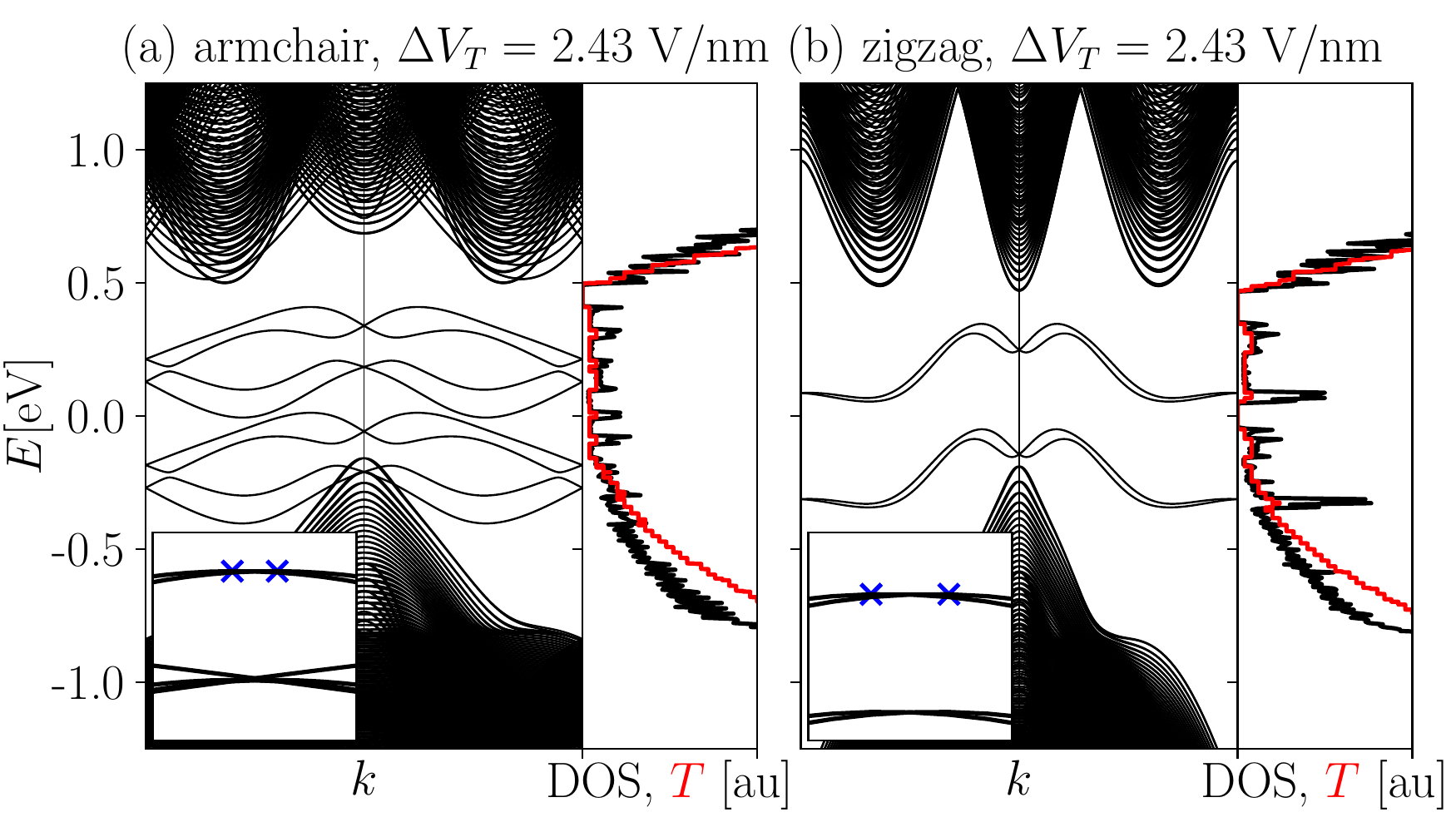}
\includegraphics[width=\linewidth]{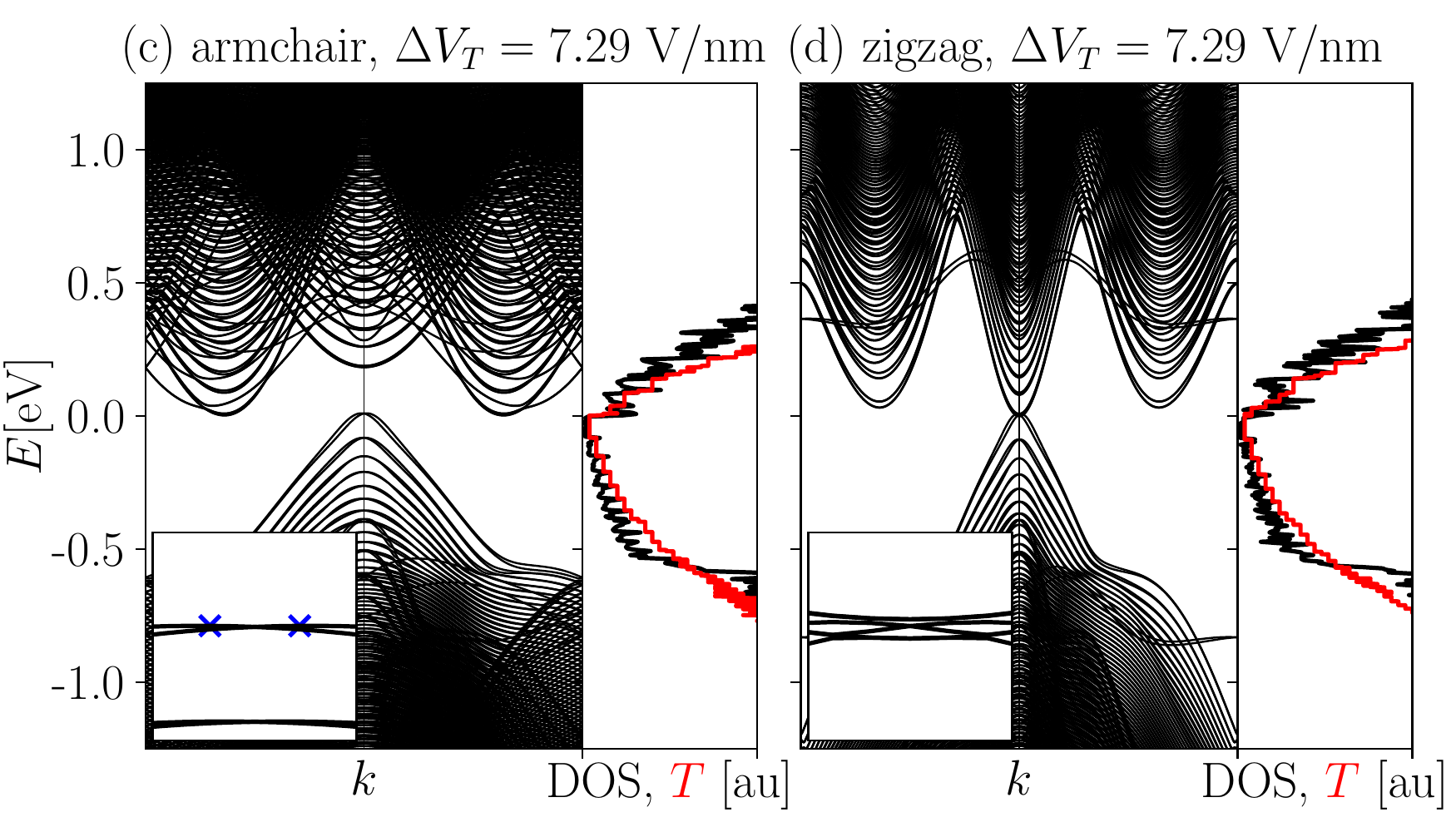}
\caption{Top: band structure, DOS and transmission for (a) an armchair ribbon of width 41 nm with $\Delta V_P= 2.43$ V/nm; (b) a zigzag ribbon of width 36 nm with $\Delta V_P= 2.43$ V/nm; (c) an armchair ribbon of width 41 nm with $\Delta V_P= 7.29$ V/nm and (d) a zigzag ribbon of width 36 nm with $\Delta V_P= 7.29$ V/nm. Insets: close up of the edge of the valence band, with blue crosses to indicate the maxima.}
\label{Fig:Transverse-DOS}
\end{figure}

\begin{figure}[t]
\centering
\includegraphics[width=0.49\linewidth]{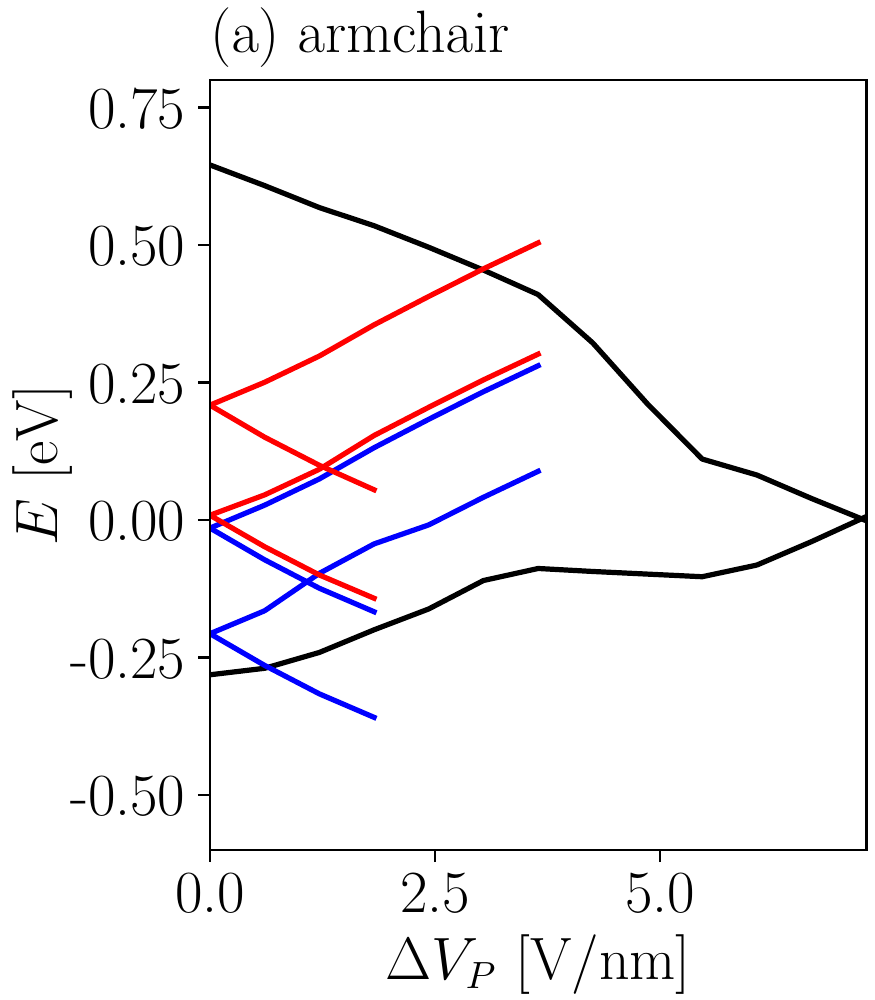}
\includegraphics[width=0.49\linewidth]{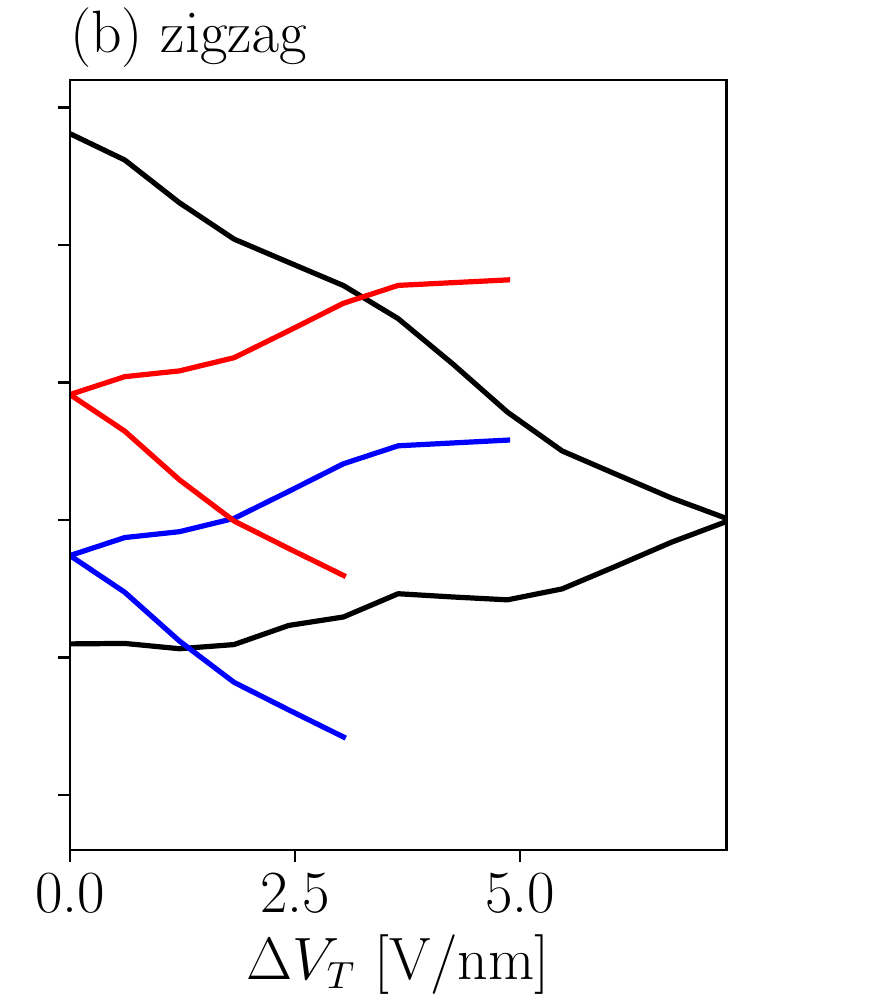}
\caption{Band edges as function of $V_T$ for (a) an armchair ribbon of width 41 nm and (b) a zigzag ribbon of width 36 nm. 
Black lines correspond to the conduction and valence band edges, and red (blue) corresponds to the maxima (minima) of the edge states. The curves corresponding to edge states are cut when such states can no longer be distinguished from the conduction or valence band states in the band structure.}
\label{Fig:Transverse-Gap}
\end{figure}

Another possibility to tune the band structure is to apply an in-plane electric field. This way we create a transverse bias potential along the ribbon. To account for a transverse bias $\Delta V_T$, we set in Hamiltonian (\ref{Eq:Hamiltonian_Real}) the on-site energy
\begin{equation}
	\epsilon_{mi\sigma} = \Delta V_T \times x_i,
\end{equation}
where $x_i$ is the coordinate of site $i$ in the transverse direction. Our results for two different values of bias field are given in Fig. \ref{Fig:Transverse-DOS}. First, we observe a band gap reduction for both types of edge termination. The evolution of the valence and conduction band edges, as well as the position of the extrema of the edge states, are shown in Fig. \ref{Fig:Transverse-Gap}. Notice that the color lines, corresponding to the edge states, are cut when such states can no longer be distinguished from the conduction or valence band states in the band structure. Similarly as in the case of an out-of-plane field, inversion symmetry is broken along the ribbon, and the edge states are split (see insets of Fig. \ref{Fig:Transverse-DOS}).

\begin{figure}[t]
\centering
\includegraphics[width=0.49\linewidth]{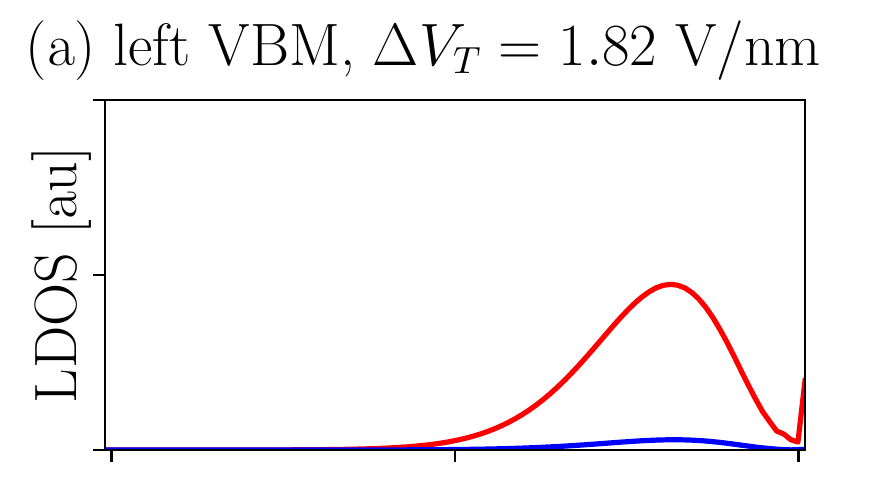}
\includegraphics[width=0.49\linewidth]{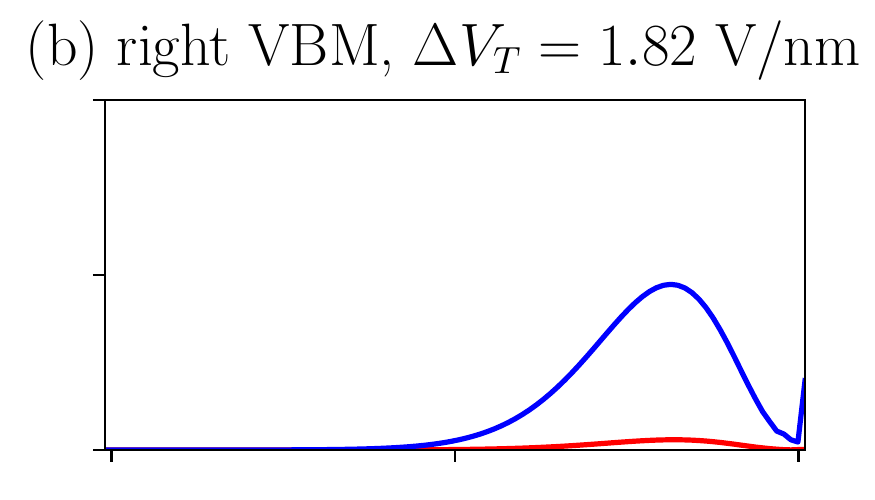} \\
\includegraphics[width=0.49\linewidth]{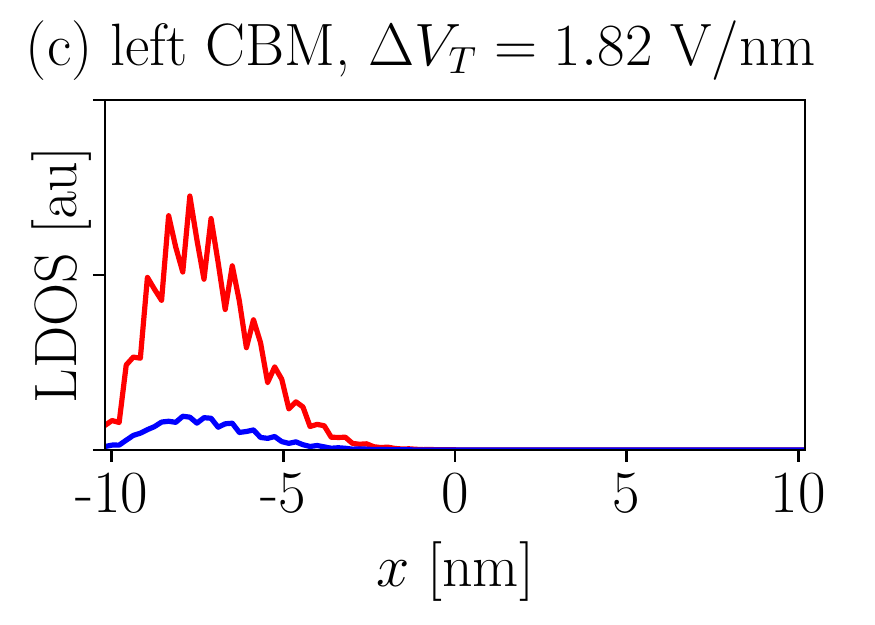}
\includegraphics[width=0.49\linewidth]{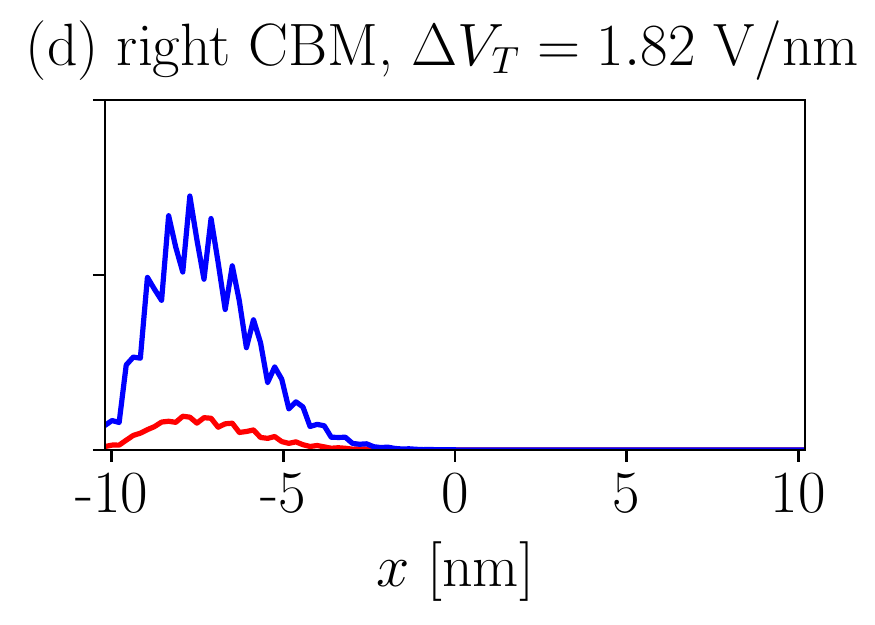}
\caption{LDOS of the valence band maxima for an armchair ribbon of width 20 nm, under transverse bias. The red and blue lines correspond to spin up and down, respectively.}
\label{Fig:Transverse-LDOS}
\end{figure}

The main difference with respect to the case of out-of-plane bias is that the edges of the valence and conduction bands correspond to states that are located at the edges of the ribbon. This is clearly seen in the LDOS calculations (Fig. \ref{Fig:Transverse-LDOS}). While the states that form the valence band edge are placed at the right edge (see Fig. \ref{Fig:Transverse-LDOS} (a) and (b)), the conduction band edge is located at the left edge of the ribbon, as it can be seen in Fig. \ref{Fig:Transverse-LDOS} (c) and (d). 

As the sites on the edges gain an on-site potential $\pm \Delta V_T \frac{W}{2}$, the edge states on one side of the ribbon move up in energy, while on the other side they move down. Otherwise, the shape of the edge states stays the same.

These results are similar to those obtained for a black phosphorus nanoribbon in the presence of a transverse electric field \cite{Taghizadeh_PRB_2015}.  The local separation between the conduction and valence bands states can be quantified by calculating the polarization \cite{yuan2010modeling}
%\begin{equation}
$
	P = e \sum_{mi\sigma} \mathbf{r}_i c^{\dagger}_{mi\sigma} c^{\phantom{\dagger}}_{mi\sigma},
$
%\end{equation}
which yields $\langle P \rangle_{\text{VBM}}=5.61e$  $\text{nm}$ and $\langle P \rangle_{\text{CBM}}=-7.25e$  $\text{nm}$ for this configuration.

Moreover, applying a transverse bias causes the valence band to split, lifting the degeneracy of the predominantly spin up and spin down states at the VBM. This is due to the fact that Rashba coupling is also present in this case (see insets in Fig. \ref{Fig:Transverse-DOS}), which leads to different spin polarization of the two extrema of the valence band, represented by different color of the two VBM (panels (a) and (b) of Fig. \ref{Fig:Transverse-LDOS}). The armchair CBM also becomes spin-polarized. For the zigzag case, and for ribbons with $W<62$ nm, the CBM is at $k=0$, where there is no spin-polarization.

\section{Conclusion}\label{Sec:Conclusions}

In summary, we have studied the electronic properties of antimonene nanoribbons, in the presence of out-of-plane and in-plane electric fields, using a tight-binding model. We have shown that there is good agreement between \textit{ab initio} results and the tight-binding model. Our calculations show that antimonene nanoribbons are semiconducting in their bulk, i.e., not taking edge states into account. We have found that, contrary to phosphorene, both kinds of termination, zigzag and armchair, present edge states inside the gap. Under the application of external bias fields, we have demonstrated that the gap can be enhanced by applying an out-of-plane bias. Under a transverse in-plane electric field, the gap decreases. Furthermore, a transverse bias leads to spatial separation between the states forming the edges of the valence and conduction bands. Both types of bias cause valence band splitting, due to Rashba coupling induced by lack of inversion symmetry. Such splitting is accompanied by a different spin polarization of the two mini-valleys at both sides of $\Gamma$ point.

 \acknowledgments 
 
JY acknowledges financial support from MOST 2017YFA0303404, NSAF U1530401 and computational resources from the Beijing Computational Science Research Center. MIK acknowledges financial support from the European Research Council Advanced Grant program (Contract No. 338957). SY acknowledges financial support from Thousand Young Talent Plan (China). RR acknowledges financial support from the Spanish MINECO through Ram\'on y Cajal program,  Grants No. RYC-2016-20663 and Grant No. FIS2014- 58445-JIN.

\bibliography{sb_modelNotes}

\end{document}